# A new metric for robustness with respect to virus spread

R.E. Kooij[1,2,3*], P. Schumm[3], C. Scoglio[3]

[1] Delft University of Technology, EEMCS, the Netherlands
[2] TNO ICT, the Netherlands
[3] Kansas State University, EECE, U.S.A.
robert.kooij@tno.nl, {pbschumm, caterina}@ksu.edu

**Abstract.** The robustness of a network is depending on the type of attack we are considering. In this paper we focus on the spread of viruses on networks. It is common practice to use the epidemic threshold as a measure for robustness. Because the epidemic threshold is inversely proportional to the largest eigenvalue of the adjacency matrix, it seems easy to compare the robustness of two networks. We will show in this paper that the comparison of the robustness with respect to virus spread for two networks actually depends on the value of the effective spreading rate $\tau$. For this reason we propose a new metric, the viral conductance, which takes into account the complete range of values $\tau$ can obtain. In this paper we determine the viral conductance of regular graphs, complete bi-partite graphs and a number of realistic networks.

**Keywords:** Robustness, virus spread, epidemic threshold, viral conductance.

## 1 Introduction

Our daily activities rely increasingly on complex networks. The power grid, the Internet, and transportation networks are examples of complex networks. In contrast to simple networks, such as regular or Erdös-Rényi random graphs [7], complex networks are characterized by a large number of vertices (from hundreds of thousands to billions of nodes), a low density of links, clustering effects, and power-law node-degree distribution [1], [20]. Being so large, complex networks are often controlled in a decentralized way and show properties of self-organization. However, even if decentralization and self-organization theoretically reduce the risk of failure, complex networks can experience disruptive and massive failure.

As an example of massive attacks, in 2001, Code Red, a computer virus that incapacitated numerous networks, resulted in a global loss of 2.6 billion US dollars. In 2004, the Sassar virus caused Delta airlines to cancel 40 transatlantic flights in addition to halting trains in Australia. Additionally, the US General Accounting Office estimated 250,000 annual attacks on Department Of Defense networks.

---

[*] Corresponding author: R.E. Kooij, TNO ICT, Brassersplein 2, 2612 CT Delft, the Netherlands

Objectives of such attacks range from theft, modification, and destruction of data to dismantling of entire networks. In another example concerning the power grid, the Northeastern and Midwestern United States, and Ontario, Canada suffered a massive widespread power outage on August 14, 2003. Since our daily routines would cease if the technological information infrastructure disintegrates, thus, it becomes crucial to maintain the highest levels of robustness in complex networks.

Therefore, the first step is to assess the robustness of networks. Obviously the robustness of a network is depending on the type of attack we are considering. In this paper we will focus on the spread of viruses on computer networks.

The Susceptible-Infected-Susceptible (SIS) infection model, which arose in mathematical biology, is often used to model the spread of computer viruses [10], [9], [14], epidemic algorithms for information dissemination in unreliable distributed systems like P2P and ad-hoc networks [3], [8], and propagation of faults and failures in networks like BGP [4]. The SIS model assumes that a node in the network is in one of two states: infected and therefore infectious, or healthy and therefore susceptible to infection. The SIS model usually assumes instantaneous state transitions. Thus, as soon as a node becomes infected, it becomes infectious and likewise, as soon as a node is cured it is susceptible to re-infection. There are many models that consider more aspects like incubation periods, variable infection rate, a curing process that takes a certain amount of time and so on [6], [10], [19]. In epidemiological theory, many authors refer to an epidemic threshold $\tau_c$, see for instance [6], [2], [10] and [14]. If it is assumed that the infection rate along each link is β while the curing rate for each node is δ then the effective spreading rate of the virus can be defined as $\tau = \beta/\delta$. The epidemic threshold can be defined as follows: for effective spreading rates below $\tau_c$ the virus contamination in the network dies out - the mean epidemic lifetime is of order log n, while for effective spreading rates above $\tau_c$ the virus is prevalent, i.e. a persisting fraction of nodes remains infected with the mean epidemic lifetime [9] of the order $\exp(n^\alpha)$. In the case of persistence we will refer to the prevailing state as a metastable state or steady state. It was shown in [18] and [9] that $\tau_c = 1/\rho(A)$ where $\rho(A)$ denotes the spectral radius of the adjacency matrix A of the graph. Recently, the epidemic threshold formula has also been verified by using the N-intertwined model [17], which consists of a pair of interacting continuous Markov chains.

It is common practice to use the epidemic threshold as a measure for robustness: the larger the epidemic, the more robust a network is against the spread of a virus, see [11]. Because the epidemic threshold is inversely proportional to the largest eigenvalue of the adjacency matrix, it seems easy to compare the robustness of two networks. We will show in this paper that the comparison of the robustness with respect to virus spread for two networks is not so straightforward. To be more precise, we will show that the comparison of networks depends on the actual value of the effective spreading rate τ. For this reason we propose a new metric, the viral conductance, which takes into account the complete range of values τ can obtain.

The rest of this paper is organized as follows. In Section 2 we consider the spread of viruses on regular and complete bi-partite graphs and show the need for a new metric for robustness with respect to virus spread. We propose this new metric, the viral conductance, in Section 3. In Section 4 we suggest a heuristic for the computation of the viral conductance. We determine the viral conductance for some realistic networks in Section 5. The main conclusions are summarized in Section 6.

## 2  Virus spread on regular and complete bi-partite graphs

In this section we will compare the fraction of infected nodes for two example networks and show that the value of the effective spreading rate $\tau$ determines for which network this fraction is higher.  The example networks belong to the class of regular and complete bi-partite graphs, respectively.

### 2.1  Virus spread on regular graphs

In this subsection we discuss the spread of viruses over a simpler network, i.e. the connected regular graph. This model is based on a classical result by Kephart and White [10] for SIS models.

We consider a connected graph on N nodes where every node has degree k. We denote the number of infected nodes in the population at time t by I(t). If the population N is sufficiently large, we can convert I(t) to i(t)=I(t)/N, a continuous quantity representing the fraction of infected nodes. Now the rate at which the fraction of infected nodes changes is due to two processes: susceptible nodes becoming infected and infected nodes being cured. Obviously, the cure rate for a fraction i of infected nodes is $\delta i$. The rate at which the fraction i grows is proportional the fraction of susceptible nodes, i.e. 1-i. For every susceptible node the rate of infection is the product of the infection rate per node ($\beta$), the degree of the node (k) and the probability that on a given link the susceptible node connects to an infected node (i).

Therefore we obtain the following differential equation describing the time evolution of i(t):

$$\frac{di}{dt} = \beta k i (1-i) - \delta i. \qquad (1)$$

The steady state solution $i_\infty$ of Eq. (1) satisfies

$$i_\infty = \frac{\beta k - \delta}{\beta k} = 1 - \frac{1}{k\tau}. \qquad (2)$$

Because an epidemic state only exists if $i_\infty > 0$, we conclude that the epidemic threshold satisfies

$$\tau_c = \frac{1}{k}. \qquad (3)$$

Because for k-regular graphs the spectral radius of the adjacency matrix is equal to k, see [5], Eq. (3) is in line with the result by [18].

### 2.2 Virus spread on complete bi-partite graphs

In this subsection we will consider complete bi-partite graphs. A complete bi-partite graph $K_{M,N}$ consists of two disjoint sets $S_1$ and $S_2$ containing respectively M and N nodes, such that all nodes in $S_1$ are connected to all nodes in $S_2$, while within each set no connections occur. Fig. 1 gives an example of a complete bi-partite graph on 10 nodes.

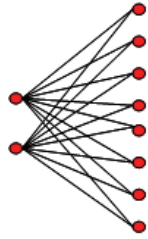

**Fig. 1.** Complete bi-partite graph $K_{2,8}$

Notice that (core) telecommunication networks often can be modeled as a complete bi-partite topology. For instance, the so-called double-star topology (i.e. $K_{M,N}$ with M = 2) is quite commonly used because it offers a high level of robustness against link failures. For example, the Amsterdam Internet Exchange (see www.ams-ix.net), one of the largest public Internet exchanges in the world, uses this topology to connect its four locations in Amsterdam to two high-density Ethernet switches. Sensor networks are also often designed as complete bi-partite graphs.

In [12] a model for virus spreading on the complete bi-partite graph $K_{M,N}$ was presented. Using differential equations and two-state Markov processes it was shown in [12] that, above the epidemic threshold $\tau_c = \frac{1}{\sqrt{MN}}$, the fraction of infected nodes for $K_{M,N}$ satisfies

$$y_\infty = \frac{(MN\tau^2 - 1)((M+N)\tau + 2)}{\tau(M+N)(M\tau + 1)(N\tau + 1)}. \qquad (4)$$

It is easy to verify that for the case M = N, Eq. (4) reduces to Eq. (2), with k = N.

### 2.3  Comparing the fraction of infected nodes for two graphs

In this subsection we consider two networks on 10 nodes, the Petersen graph (see Fig. 2) and $K_{2,8}$. Note that the Petersen graph is a regular graph where every node has 3 neighbours, i.e. k = 3 in the notation of section 2.1.

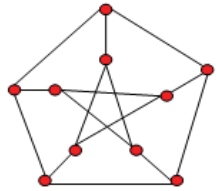

**Fig. 2.** Petersen graph

Using Eq. (2) and Eq. (4) we can compare the fraction of infected nodes for the Petersen graph and $K_{2,8}$, see Fig. 3.

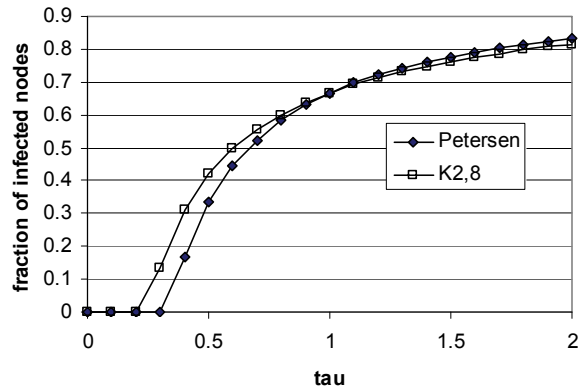

**Fig. 3.** Fraction of infected nodes for Petersen graph and $K_{2,8}$

If we only look at the epidemic threshold, we see that the Petersen graph outperforms $K_{2,8}$. However, for τ sufficiently large, $K_{2,8}$ performs better because then the fraction of infected nodes is lower than for the regular graph. Note that this effect, that regular graphs have a higher fraction of infected nodes than non-regular graphs for large values of τ, was already observed in [16].

This section shows that in order to compare for two networks the robustness with respect to virus spread, it does not suffice only to look at the epidemic threshold.

## 3  Viral conductance

In this section we propose a new metric for robustness with respect to virus spread that takes into account the complete range of τ values.

A natural way to take all values of τ into account is by considering the area under the curve that gives the fraction of infected nodes. However, because this will lead to divergent integrals, from now on, instead of considering the effective spreading rate τ we look at the reciprocal of τ, that is the effective spreading rate $s = \delta/\beta$.

We are interested in $y_\infty(s)$, the fraction of infected nodes in steady state, as a function of the effective curing rate. Note that the behaviour of $y_\infty(s)$ around $s = 0$ reflects the behaviour of the original system for $\tau \to \infty$.

We are now in the position to suggest a new robustness measure with respect to virus spread that takes into account all values of τ, and hence s.

**Definition**. The viral conductance V of a network G is given by

$$V(G) = \int_0^\infty y_\infty(s)\,ds, \tag{5}$$

where $y_\infty(s)$ denotes the fraction of infected nodes in steady state and $s = \delta/\beta$.

We will now state some theorems for the viral conductance V(G).

**Theorem 1**: For regular graphs $H_k$, where every node has k neighbours, it holds that $V(H_k) = k/2$.

Proof. This follows directly from Eq. (2) and Eq. (5).

**Theorem 2**: For complete bi-partite graphs $K_{M,N}$, it holds that $V(K_{M,N}) =$

$$\frac{(M+N)\sqrt{MN} - MN + (M-N)(N\ln(M+\sqrt{MN}) - M\ln(N+\sqrt{MN}) + M\ln M - N\ln N)}{M+N}.$$

Proof. This follows from applying Eq. (5) to Eq. (4).

## 4 A heuristic for the viral conductance

For general networks we cannot compute the fraction of infected nodes $y_\infty(s)$, and hence the viral conductance, explicitly. Therefore, in this section we propose a heuristic for the computation of the viral conductance for general networks.

We start with listing some properties of the fraction of infected nodes $y_\infty(s)$.

**Lemma 1** For any connected graph G let A denote its adjacency matrix and $\rho(A)$ the largest eigenvalue of A. Then $y_\infty(\rho(A))=0$.

Proof. This is just the threshold theorem, see e.g. [17].

**Lemma 2** Consider a connected graph on N nodes and denote the degree of node i by $d_i$. Then $y_\infty(s) = 1 - s\frac{1}{N}\sum_{i=1}^{N}\frac{1}{d_i} + O(s^2)$.

Proof. This follows from Section IV in [17].

**Conjecture** For any connected graph G on N nodes, let A denote its adjacency matrix, $\rho(A)$ the largest eigenvalue of A and $d_i$ the degree of node i. Then
$$\lim_{s\uparrow\rho}\frac{dy_\infty}{ds} = -\frac{E[d_i]}{\rho^2}.$$

Motivation. The conjecture is true for regular and complete bi-partite graphs and also is supported by numerical evidence.

We will use the above lemmas and conjecture to construct a heuristic for estimating the new robustness metric V.

We first approximate the curve $y_\infty(s)$ by two straight lines, $y_{\infty 1}(s)$ and $y_{\infty 2}(s)$ such that $y_{\infty 1}(s)$ is the linearization of $y_\infty(s)$ at $s = 0$ and $y_{\infty 2}(s)$ is the linearization of $y_\infty(s)$ at $s = \rho$.

From Lemma 2 and the conjecture it follows that

$$y_{\infty 1}(s) = 1 - \sigma s, \quad y_{\infty 2}(s) = \frac{E[d_i]}{\rho} - \frac{E[d_i]}{\rho^2}s$$

where $\sigma = \frac{1}{N}\sum_{i=1}^{N}\frac{1}{d_i}$. The lines $y_{\infty 1}(s)$ and $y_{\infty 2}(s)$ intersect at

$$s^* = \frac{\rho(\rho - E[d_i])}{\sigma\rho^2 - E[d_i]}.$$

Hence we obtain for this heuristic

$$V_{PL} = \int_0^{s^*} y_{\infty 1}(s)ds + \int_{s^*}^{\rho} y_{\infty 2}(s)ds = \frac{\rho((\sigma\rho - 2)E[d_i] + \rho)}{2(\sigma\rho^2 - E[d_i])}..$$

Next we approximate the curve $y_\infty(s)$ by a non-linear curve $y_{\infty NL}(s)$ of the following form $y_{\infty NL}(s) = a + bs + cs^d$ such that $y_{\infty NL}(s)$ and $y_\infty(s)$ have the same linearization, both at $s = 0$ and at $s = \rho$. A straightforward calculation shows that

$$y_{\infty NL}(s) = 1 - \sigma s + (\sigma\rho - 1)\left(\frac{s}{\rho}\right)^d,$$

where $d = \frac{\sigma\rho^2 - E[d_i]}{\rho(\sigma\rho - 1)}$.

Note that for non-regular graphs $\rho > E[d_i]$, see e.g. [5], from which it can be deduced that $d > 1$, hence the derivative of $y_{\infty NL}(s)$ at $s = 0$ is finite.

Hence we obtain for this heuristic

$$V_{NL} = \int_0^{\rho} y_{\infty NL}(s)ds = \frac{\rho((\sigma\rho - 2)E[d_i] + \sigma\rho^2)}{2(2\sigma\rho^2 - E[d_i] - \rho)}..$$

Because $V_{NL}$ always seems to lead to an overestimation of V, while $V_{PL}$ underestimates V, as the final heuristic $V_H$ for the viral conductance we propose:

$$V_H = \frac{V_{PL} + V_{NL}}{2}.$$

We will now validate the heuristic for the complete bi-partite graph $K_{M,N}$, for which we can determine the new robustness measure V explicitly, see Theorem 2.

The results are given in Table 1.

**Table 1:** Comparison viral conductance V with heuristic $V_H$ for $K_{M,N}$

| M | N | V | $V_H$ | Relative error |
|---|---|---|---|---|
| 10 | 90 | 11.38 | 11.22 | -1.42% |
| 30 | 70 | 21.85 | 21.77 | -0.32% |
| 50 | 50 | 25 | 25 | 0% |
| 10 | 990 | 20.14 | 21.95 | 8.24% |
| 100 | 900 | 113.77 | 112.18 | -1.42% |
| 250 | 750 | 200.24 | 199.07 | -0.59% |

Apart from the case $K_{10,990}$, the heuristic $V_H$ performs very well.

## 5 Viral conductance for realistic networks

In this section we will determine the viral conductance for a number of real-life networks and for some toy networks that are commonly used to model realistic networks.

### 5.1 Considered networks

The following networks are considered in this section: the Abilene backbone network, a scale free network, HOT (Heuristically Optimal Topology), the Erdös-Rényi graph, the Stanley Ring, the Stanley Mesh and a 2D-lattice. The Stanley Ring and Stanley Mesh were proposed in [15] to maximize network robustness against both random and targeted attacks while minimizing the network cost.
More information on the considered networks can be found in [12], [15] and [16].

All networks contain approximately 1000 nodes. This value reflects the number of nodes in current inter-domain network design and thus provides a comparatively real-life scenario. Some of the considered networks are visualized in Fig. 4.

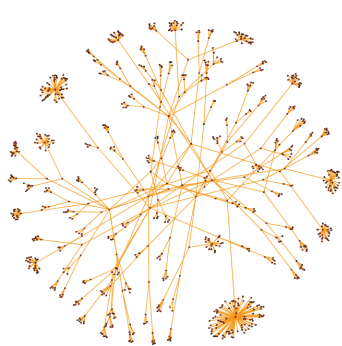 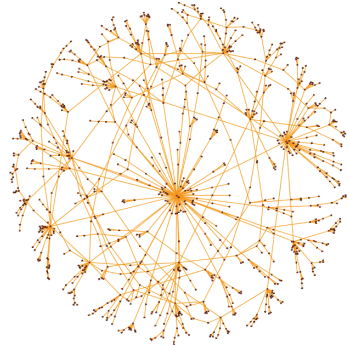

a) Abilene network           b) Scale free network

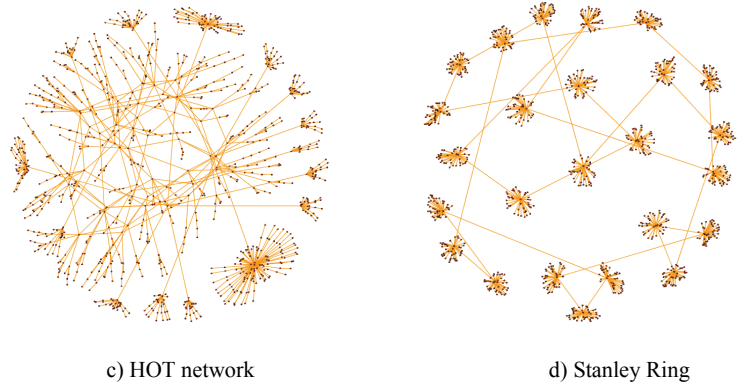

c) HOT network　　　　　　　　　d) Stanley Ring

**Fig. 4.** Visualization of some of the considered networks

### 5.2 Numerical results

Because for general networks no explicit expression for the fraction of infected nodes $y_\infty(s)$ is available, we have used numerical analysis to determine the viral conductance for the considered networks. The first step is to obtain steady state values for the number of infected nodes, for a given value of the effective curing rate s. For this we use the discrete, deterministic expression for $p_{i,t}$, the probability that node i is infected at time t, as given in [18]. By summing over all nodes and after the appropriate rescaling we obtain $y_\infty(s)$. This fraction of infected nodes is evaluated for 100 equidistant values of s, between 0 and $\rho(A)$, the spectral radius of the adjacency matrix A. Finally, the viral conductance is determined by means of a simple triangular integration method. The results are given in Table 2.

**Table 2:** Viral conductance V for realistic networks

| Network | N | L | <d> | $\tau_c$ | V | $V_H$ | rel. error |
|---|---|---|---|---|---|---|---|
| Abilene | 886 | 896 | 2.02 | 0.11 | 1.43 | 2.18 | 53% |
| Scale free | 1000 | 1049 | 2.10 | 0.10 | 1.49 | 2.26 | 52% |
| HOT | 1000 | 1049 | 2.10 | 0.11 | 1.46 | 2.23 | 53% |
| Erdös-Rényi | 1000 | 2009 | 4.02 | 0.19 | 2.20 | 2.24 | 2% |
| Stanley Ring | 1000 | 1000 | 2.00 | 0.14 | 1.79 | 1.84 | 3% |
| Stanley Mesh | 1000 | 1275 | 2.55 | 0.04 | 2.81 | 4.41 | 57% |
| 2D-lattice | 900 | 1740 | 3.87 | 0.25 | 2.00 | 1.96 | -2% |

In Table 2 N denotes the number of nodes, L the number of links and <d> the average nodal degree. $V_H$ denotes the viral conductance according to the heuristic proposed in Section 4. Several conclusions can be drawn from Table 2. We confine ourselves to mention just a few:

- Of the considered networks Abilene has the lowest viral conductance; hence it is the most robust with respect to virus spread.
- The threshold of the Erdös-Rényi graph is almost twice as high as that of Abilene, yet its viral conductance is about 50% higher.
- The heuristic $V_H$ is very accurate for the Erdös-Rényi graph, the Stanley Ring and the 2D-lattice. For the other networks it leads to an overestimation of the viral conductance in the order of 50%.
- Rewiring the links of the Stanley Ring could lead to a reduction of the viral conductance of about 44%.

The last conclusion follows from the fact that every ring topology has a viral conductance of 1, see Theorem 1.

## 6 Conclusions

In this paper we have proposed the viral conductance as a new metric for robustness with respect to virus spread in networks. The viral conductance takes the complete range of values the effective spreading rate can attain into account. We have given an explicit expression for the viral conductance in case of regular and complete bi-partite graphs. For general networks we have proposed a heuristic. Next we have determined the viral conductance for a number of realistic networks, by means of numerical computation.

The main conclusions are the following:
- of the considered networks Abilene has the lowest viral conductance; hence it is the most robust with respect to virus spread;
- the heuristic $V_H$ is very accurate for the Erdös-Rényi graph, the Stanley Ring and the 2D-lattice. For the other networks it leads to an overestimation of the viral conductance in the order of 50%.

The following issues will be subject of our future work:
- design of topologies, with given number of nodes and links, which minimize the viral conductance;
- construction of a more accurate heuristic for the computation of the viral conductance;
- computation of the viral conductance for networks of a larger scale.

## 7 Acknowledgement

This research was partially supported by the Netherlands Organization for Scientific Research (NWO) under project number 643.000.503 and by the National Science Foundation (NSF) under award number 0841112.